\documentclass[10pt, conference, letterpaper]{IEEEtran}
\IEEEoverridecommandlockouts

\usepackage{cite}
\usepackage{amsmath,amssymb,amsfonts}
\usepackage{amsthm}
\usepackage{mathtools}
\usepackage{enumitem}
\usepackage{algorithmic}
\usepackage{graphicx}
\usepackage{wrapfig}
\usepackage{textcomp}
\usepackage[table]{xcolor}
\usepackage{tcolorbox}
\def\BibTeX{{\rm B\kern-.05em{\sc i\kern-.025em b}\kern-.08em
    T\kern-.1667em\lower.7ex\hbox{E}\kern-.125emX}}

\usepackage{tabularx}
\usepackage{booktabs}
\usepackage{threeparttable}
\usepackage{makecell}
\usepackage{pifont}
\usepackage{hyperref}
\usepackage{tikz}
\usepackage{inconsolata}
\usepackage[linesnumbered,ruled,vlined]{algorithm2e}
\SetCommentSty{\fontsize{5}{5}\selectfont} 
\SetKwComment{tcc}{$\triangleright\,$}{}

\SetCommentSty{mycommfont}
\SetKwInput{KwData}{Require}
\SetKwInput{KwResult}{Result}

\definecolor{mygreen}{RGB}{11,141,10}
\definecolor{myred}{RGB}{223,68,52}
\definecolor{myblue}{RGB}{70,130,180}
\definecolor{mydeepblue}{RGB}{65,105,225}
\definecolor{myviolet}{RGB}{97,0,138}
\definecolor{myburgundy}{RGB}{110,10,30}
\definecolor{myblue2}{RGB}{0,105,148}
\definecolor{grayhighlight}{RGB}{250,250,227}

\hypersetup{
colorlinks=true,
linkcolor=black,
anchorcolor=blue,
citecolor=black,
urlcolor=black,
}

\newcommand{\linear}[2]{\textcolor{blue}{\langle}#1\textcolor{blue}{\cdot}#2\textcolor{blue}{\rangle}}
\usepackage{stmaryrd}
\newcommand{\keyshift}[2]{\llbracket {#1} \rrbracket_{#2}}

\newtheorem{theorem}{Theorem}[section]

\newcommand{\ie}{{\em i.e.}}
\newcommand{\eg}{{\em e.g.}}
\newcommand{\aka}{{\em a.k.a.}}
\newcommand{\wrt}{{\em w.r.t.}}
\newcommand{\etc}{{\em etc.}}

\newcommand{\name}{$\mathrm{Tempo}$}
\newcommand{\enc}{{MM-obfuscation}}

\usepackage{siunitx} %

\usepackage{cryptocode}

\begin{document}

\title{\name{}: Confidentiality Preservation in Cloud-Based Neural Network Training\\
}

\author{
  \IEEEauthorblockN{Rongwu Xu and
  Zhixuan Fang}
  \IEEEauthorblockA{IIIS, Tsinghua University}

  \IEEEauthorblockA{\texttt{xrw22@mails.tsinghua.edu.cn}, \texttt{zfang@mail.tsinghua.edu.cn}}
}

\maketitle

\begin{abstract}
Cloud deep learning platforms provide cost-effective deep neural network (DNN) training for customers who lack computation resources. However, cloud systems are often untrustworthy and vulnerable to attackers, leading to growing concerns about model privacy. Recently, researchers have sought to protect data privacy in deep learning by leveraging CPU trusted execution environments (TEEs), which minimize the use of cryptography, but existing works failed to simultaneously utilize the computational resources of GPUs to assist in training and prevent model leakage. This paper presents \name{}, the first cloud-based deep learning system that cooperates with TEE and distributed GPUs for efficient DNN training with model confidentiality preserved. To tackle the challenge of preserving privacy while offloading linear algebraic operations from TEE to GPUs for efficient batch computation, we introduce a customized permutation-based obfuscation algorithm to blind both inputs and model parameters. An optimization mechanism that reduces encryption operations is proposed for faster weight updates during backpropagation to speed up training. We implement \name{} and evaluate it with both training and inference for two prevalent DNNs. Empirical results indicate that \name{} outperforms baselines and offers sufficient privacy protection.
\end{abstract}

\section{Introduction}
\label{sec:intro}

With the rapid advancement of deep learning (DL)~\cite{lecun2015deep} and cloud computing services~\cite{googleaiplatform,microsoftMLaaS}, clients are motivated to \emph{outsource} computationally intensive DL tasks to high-performance cloud platforms, especially when local devices lack sufficient resources.
This trending ``Cloud DL'' approach, \aka, Machine Learning as a Service (MLaaS)~\cite{ribeiro2015mlaas}, is actively supported by major cloud service providers (CSPs) like Google Cloud~\cite{googleaiplatform} and Microsoft Azure~\cite{microsoftMLaaS}.
Despite its convenience, the MLaaS paradigm inevitably introduces concerns regarding security and data privacy~\cite{ghodsi2017safetynets,hunt2020telekine}, as CSPs can be untrustworthy, self-interested, and possibly malicious~\cite{wang2010privacy}.

\noindent \textbf{Motivation.} 
Consider a motivating example, where an animation company wishes to employ its own dataset to fine-tune a pre-trained model provided by OpenAI for AIGC~\cite{ramesh2022hierarchical} tasks. To manage costs, the company expects to train the model on a CSP. Unfortunately, direct outsourcing the job to the cloud is not \emph{secure} for two privacy (\aka, confidentiality) concerns:
\begin{itemize}[leftmargin=*]
    \item \emph{Input privacy}: The company does not want to expose its training dataset to any third party, including the CSP.
    \item \emph{Model privacy}: As the size of models grows exponentially, the computational resources and data needed for training also increase drastically. The resulting model, acquired through an expensive training process, holds substantial commercial value~\cite{tramer2016stealing,jia2018preserving} (\eg, SOTA LLMs like ChatGPT and GPT-4 are proprietary and not yet open-source). Consequently, the trained model, as an \emph{intellectual property (IP)}, should not be leaked to any third party.
\end{itemize}

\begin{table}

\centering\fontsize{7}{7.75}\selectfont

\newcommand{\cmark}{\textcolor{mygreen}{\ding{51}}}%
\newcommand{\xmark}{\textcolor{darkgray}{\ding{55}}}%
\newcommand*\emptycirc[1][0.7ex]{\tikz\draw (0,0) circle (#1);} 
\newcommand*\halfcirc[1][0.7ex]{%
  \begin{tikzpicture}
  \draw[fill=darkgray] (0,0)-- (90:#1) arc (90:270:#1) -- cycle ;
  \draw (0,0) circle (#1);
  \end{tikzpicture}}
\newcommand*\fullcirc[1][0.7ex]{\tikz\fill (0,0) circle (#1);} 

\centering

\caption{Comparison of \name{} and notable related systems.}
\vspace{-0.25cm}
\setlength\tabcolsep{0cm}
\begin{threeparttable}
\begin{tabularx}{\linewidth}{c*7{>{\centering\arraybackslash}X}}
\toprule
\textbf{System} &Model Privacy &Input Privacy &GPU \newline Employment &Inference &Training &Distributed\\
\midrule
Slalom~\cite{tramer2018slalom} & &\cmark &\cmark &\cmark & &\\
MLCapsule~\cite{hanzlik2021mlcapsule} &\cmark &\cmark & &\cmark & &\\
DarKnight~\cite{hashemi2021darknight} & &\cmark &\cmark &\cmark &\cmark &\halfcirc\tnote{1}\\
SOTER~\cite{shen2022soter} &\cmark &\cmark &\cmark &\cmark & &\\
Shadownet~\cite{sun2022shadownet} &\cmark &\cmark &\cmark &\cmark & &\\
\rowcolor{gray!20}\name{} (Ours) &\cmark &\cmark &\cmark &\cmark &\cmark &\cmark\\
\bottomrule
\end{tabularx}
\end{threeparttable}
\footnotesize
\begin{tablenotes}
    \fontsize{6}{8}\selectfont
    \item{1.} DarKnight supports multiple GPUs, but on the same server.
\end{tablenotes}
\label{tab:comparision}
\end{table}

\noindent \textbf{Research status and significance.}
The quest for privacy-preserving cloud DL has spurred considerable research endeavors~\cite{shokri2015privacy, gilad2016cryptonets, mohassel2017secureml}. 
Recently exploration into Trusted Execution Environments (TEEs)~\cite{tramer2018slalom,zhang2021citadel}, which are widely used to uphold code and data confidentiality~\cite{alves2004trustzone,intel2015intel}---has emerged as a promising avenue to safeguard DL privacy systematically.
Utilizing a server equipped with TEE support~\cite{li2023survey} to conduct the DL task entirely within the TEE (dubbed the ``\emph{pure TEE}'' solution) seems technically viable for safeguarding both the model and data.
However, mainstream TEEs often lack hardware accelerators like GPUs, leading to impractical training overheads that are reported to be orders of magnitude slower than direct GPU utilization~\cite{ng2021goten,hashemi2021darknight}.
To mitigate this, existing arts \emph{offload} computationally intensive algebraic computations from the TEE to co-allocated GPUs for efficiency.
To guarantee strong privacy, both the computation on GPUs and the communication among devices must be carefully blinded or encrypted.
Several advancements have focused on protecting client input privacy (\eg, \cite{tramer2018slalom,hashemi2021darknight,ng2021goten,niu20213legrace,sun2022shadownet}) by collaborating between TEE and GPUs.
\textbf{However, there remains a dire scarcity of cloud DL frameworks for model privacy protection}. Existing model privacy-preserving systems such as \cite{hou2021model,shen2022soter}, are predominantly limited to inference tasks. 
Notably, at the time of paper writing, it occurs to the authors that no MLaaS platform apprears to offer support for both TEE and GPU on the same server (\eg, Microsoft Azure\footnote{\url{https://learn.microsoft.com/en-us/answers/questions/1187799/which-azure-vm-image-support-both-intel-sgx-and-nv}}).

\begin{figure}
  \centering
  \includegraphics[width=\linewidth]{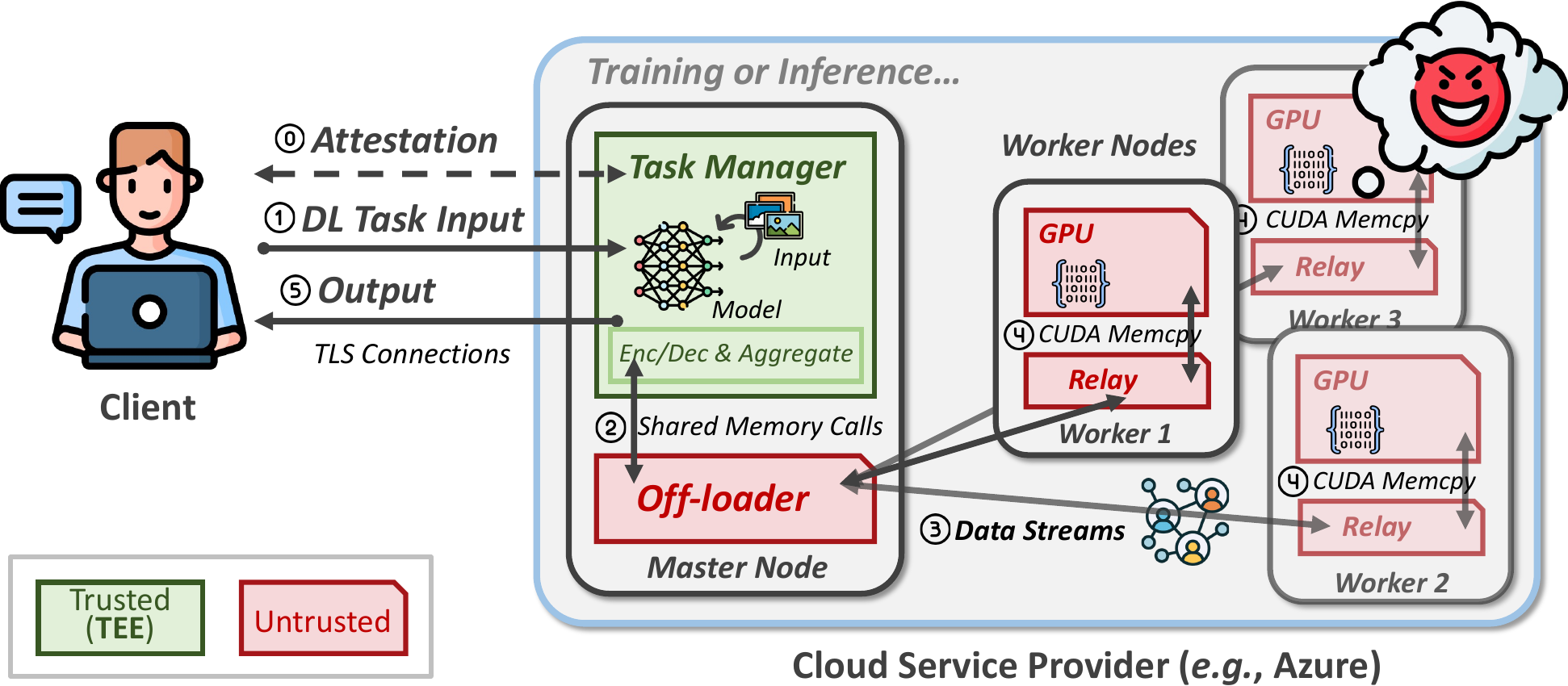}
  \vspace{-0.6cm}
  \caption{Overview of the system setup, and communications of \name{}.}
  \label{fig:system}
\end{figure}

In this paper, we propose~\name{}, where the server furnished with TEE acts as the master and collaborates with workers equipped with GPU to accomplish the DL task.
We provide an overview of how \name{} works at a high level in~\autoref{fig:system}.
\name{} is the first efficient \textbf{\underline{TE}}E-assisted \textbf{\underline{m}}odel \textbf{\underline{p}}rivacy-preserving cl\textbf{\underline{o}}ud DNN \emph{training} system. 
\autoref{tab:comparision} lists a brief comparison of \name{} with related systems.

\noindent \textbf{Challenges and solutions.} 
Two challenges arise in constructing the data-offloading scheme in \name{}:
(a) \emph{Model privacy protection}: Prior arts share the model parameters $W$ with the GPU and protect the input $X$ only. It is nontrivial to design an efficient encryption method to protect both $W$ and $X$ and guarantee the correctness of the offloaded linear computation.
(b) \emph{Efficient training}: A privacy-preserving \emph{training} system is another troublesome task since the \emph{backpropagation} involves more complicated computations: \name{} is required to keep the model weights privacy-preserved while making them available to workers for gradients calculation. 
Moreover, the constantly changing weights trigger more frequent encryption operations, which substantially increases the overhead of the system, leading to an unwanted performance degradation. 

To tackle challenge (a), we introduce a \underline{m}atrix \underline{m}ultiplication \underline{obfuscation} algorithm (\enc{}) that \emph{blinds both model weights and inputs} while providing intact training/inference accuracy.
Weights and inputs are first encrypted using a set of secret keys generated within the master's task manager (see~\autoref{fig:system}) and then streamed to workers for batched linear computation via relays, and the returned encrypted result is aggregated, decrypted, and then verified at the master to take precautions against potential active attacks (\eg, returning manipulated results) from the malicious CSP.
To address challenge (b), considering cost-efficient and time-efficient backpropagation, we optimize our training strategy by designing a novel key shift mechanism based on the proposed \enc{} algorithm to \emph{reuse computation results} from the forward pass, thus reducing the number of encryption operations by $50\%$ in a complete training epoch.

\noindent \textbf{Contributions.} We make the following contributions:
\begin{itemize}
    \item We introduce \name{}. To the best of our knowledge, this is the first distributed cloud DL training system that upholds both input and model privacy by employing a trusted master to preserve confidentiality and multiple GPU workers to enhance efficiency in training.
    \item We propose the \enc{} algorithm, a permutation-based algorithm that obfuscates both model parameters and inputs, seamlessly aligning with the functionalities of \name{}. This algorithm is optimized to reduce the encryption overhead during backpropagation, thereby accelerating DNN training.
    \item We implement \name{} on the top of Intel SGX~\cite{intel2015intel} using Graphene-SGX~\cite{tsai2017graphene} and Pytorch~\cite{pytorchcpp}. Our experiments encompass two prominent DNN architectures, ResNet~\cite{he2016deep} and Vision Transformer~\cite{dosovitskiy2020image} under diverse configurations. In comparison with pure TEE baselines, \name{} sports an average training speed-up of $4.37\times$ (local setup) and $3.95\times$ (network setup) without compromising model accuracy.
    \item We offer a comprehensive assessment of the \enc{} algorithm, scrutinizing its correctness, privacy, and cost. We launch model theft attacks on \name{}, conclusively demonstrating that our design does not provide any advantage to potential attackers.
\end{itemize}

\section{Preliminaries}
\label{sec:backgroud}

\subsection{Deep Neural Network Training}

A Deep Neural Network (DNNf) is composed of multiple sequential layers, each performing a linear or nonlinear operation. Linear operations involve trainable \emph{weights}, denotes as $\mathbf{W} =\left( \mathbf{W}_1,\mathbf{W}_2,\cdots,\mathbf{W}_L \right)$ for a DNN with $L$ layers. We denote the linear operation as $\linear{}{}$, and the non-linear operation at layer $l$ as $\sigma_l()$. 
DNN training involves a \emph{forward pass} to compute the output, followed by weights update through \emph{backpropagation} based on the loss~\cite{rumelhart1985learning}. 
This process repeats iteratively using mini-batch Stochastic Gradient Descent (SGD) until a termination criterion is met~\cite{hinton2012neural}.

\noindent\textbf{Feedforward.} 
The forward pass computes the features sequentially layer-wise, starting from layer $1$.
For layer $l$, we denote the input and output features as $\mathbf{x}_l^{i}, \mathbf{y}_l^{i}$ (here $i=1,\cdots,\mathcal{B}$ denotes the $i$th sample in the mini-batch with a size of $\mathcal{B}$).
Two neighboring layers are linked by $\mathbf{x}_{l+1}^{i} = \mathbf{y}_l^{i}$. 
The input of the DNN is $\mathbf{X}_1$ and the output is $\mathbf{Y}_L$.
Without losing of generality, we express the basic computations in forward as:
$
\mathbf{y}_l^{i} = \sigma_l(\mathbf{z}_l^{i}) = \sigma_l(\linear{\mathbf{W}_l}{\mathbf{x}_l^{i}}) 
$, in which $\mathbf{z}_l^{i}$ is the result of the linear operation.
Noticing that we can compute $\linear{\mathbf{W}_l}{\mathbf{x}_l^{i}}$ over all samples $\mathbf{X}_l = \left(\mathbf{x}_l^{1}, \mathbf{x}_l^{2}, \cdots, \mathbf{x}_l^{\mathcal{B}}\right)$ in a batch \emph{at a time}:
\begin{equation}
\label{eq:feed-forward-batch}
\mathbf{Y}_l = \sigma_l(\mathbf{Z}_l) = \sigma_l(\linear{\mathbf{W}_l}{\mathbf{X}_l}).
\end{equation}

\noindent\textbf{Backpropagation.}
The backpropagation (BP) computes the gradients recursively layer-wise in an opposite direction, starting from layer $L$.
For a sample $\mathbf{x}_1^{i}$, we use $\mathbf{y}^{i}$ to denote its label (target or ground truth).
We denote the loss function as $\mathcal{J}(\mathbf{y}_L^{i}, \mathbf{y}^{i}) = \mathcal{J}(\mathbf{W})$, which compares the predicted output with the true label, and measures how well the DNN behaves. 
Since we obtain $\mathbf{y}_L^{i}$ by feeding forward $\mathbf{x}_1^{i}$ over all layers of the DNN, therefore, $\mathcal{J}$ is also a function of all model weights $\mathbf{W}$. The basic computations in BP are:
\begin{equation}
\label{eq:back-single}
\begin{aligned}
\small
    &\mathbf{W}_l^* \gets \mathbf{W}_l - \lambda \nabla_{\mathbf{W}_l}\mathcal{J}, \\ 
    &\nabla_{\mathbf{W}_l}\mathcal{J} = \frac{\partial{\mathcal{J}}} {\partial{\mathbf{W}_l}} = \frac{1}{\left|\mathcal{B}\right|}\sum_{i=1}^{\mathcal{B}}\linear{\delta_{l}^{i}}{(\mathbf{y}_{l-1}^{i})^\top}, \\
\end{aligned}
\end{equation}
where
\begin{equation}
    \begin{cases}
    \delta_{L}^{i} =\nabla_{\mathbf{y}}\mathcal{J}(\mathbf{y}_L^{i}, \mathbf{y}^{i}) \odot \sigma'_{l}(\mathbf{z}_{L}^{i}), \\
    \delta_{l}^{i} =\linear{(\mathbf{W}_{l+1})^\top}{\delta_{l+1}^{i}}\odot\sigma'_{l}(\mathbf{z}_{l}^{i}).
    \end{cases}
\end{equation}
Here, $\mathbf{W}_l^*$ is the updated weights, $\lambda$ is the learning rate, $\delta_{l}^{i}$ is the gradient of the loss for the $i$th sample in the batch to the output of layer $l$, $\odot$ is the Hadamard product.
Again, noticing that we can compute $\linear{\delta_{l}^{i}}{(\mathbf{y}_{l-1}^{i})^\top} = \linear{\delta_{l}^{i}}{(\mathbf{x}_{l}^{i})^\top}$ and $\linear{(\mathbf{W}_{l+1})^\top}{\delta_{l+1}^{i}}$ over all samples $\mathbf{X}_l$. 
For convenience, we define $\Delta_l = \left(\delta_{l}^{1}, \delta_{l}^{2}, \cdots, \delta_{l}^{\mathcal{B}} \right)$. 
Now we rewrite~\autoref{eq:back-single} as:
\begin{equation}
\label{eq:back-batch} 
\begin{aligned}
    &\nabla_{\mathbf{W}_l}\mathcal{J} = \frac{\partial{\mathcal{J}}} {\partial{\mathbf{W}_l}} = \frac{1}{\left|\mathcal{B}\right|}\linear{\Delta_{l}}{(\mathbf{X}_{l})^\top},\\
    &\Delta_{l} =\linear{(\mathbf{W}_{l+1})^\top}{\Delta_{l+1}}\odot\sigma'_{l}(\mathbf{Z}_{l}).
\end{aligned}   
\end{equation}

\subsection{Confidential Computing using TEE and Intel SGX}

A Trusted Execution Environment (TEE) or trusted hardware, such as Arm TrustZone~\cite{alves2004trustzone}, Intel SGX~\cite{intel2015intel}, AMD SEV~\cite{amdsev}, \etc, constitutes a secure region within the CPU processor that offers hardware-assisted assurance for the privacy and integrity of a user's code and data.
Intel Software Guard eXtensions (SGX) \cite{intel2015intel,costan2016intel} is a specific TEE baked into many of Intel's x86-architecture CPUs. SGX isolates specific user-level applications (\eg, the task manager) in protected memory regions called \emph{enclaves} (TEE and enclave are interchangeable terms).
Interested readers can refer to \cite{costan2016intel} for an in-depth exposition on SGX technology.

\section{Problem Statement}
\label{sec:overview}

\subsection{Problem Setting}
We consider an outsourced cloud DL scheme illustrated in~\autoref{fig:system} involving a client and a CSP, where the CSP handles the DL task which is either training or inference.
The setup of \name{} may differ from other cloud-based ML API systems since \emph{the model is provided by the client rather than the CSP} (\ie, the client is both the data owner and model owner). 
Depending on the DL task, the input and output in~\autoref{fig:system} are varied. In the inference task, the client provides the private model with the sensitive inference input, and takes the predictions as output.
In the training task, the client provides either a model framework\footnote{Since the model framework used for training is public, it can also be provided by a (trusted) $3^{\text{rd}}$ party model provider.} or a pre-trained one and private training data, and obtains the trained private model.

\subsection{Adversarial Model} 
The CSP is \textbf{malicious}, meaning that he can arbitrarily deviate from any prescribed protocol and thus eavesdrop sensitive information through conducting active attacks (\eg, returning dishonest results) or performing lazy computation to save computing or storage costs.
We assume that SGX side-channel attacks~\cite{van2018foreshadow, murdock2020plundervolt} are beyond the scope of this paper.

\subsection{Design Goals}
\label{subsec: goals}
\name{} aims to meet the following two design goals:

\begin{enumerate}[leftmargin=*]
\item
\emph{Privacy w.r.t. Model and Input} (\autoref{th:privacy}): The adversary learns no useful information about the model weights and inputs, i.e., it would take non-polynomial time for the adversary to recover the information in practice.
\item
\emph{Verifiable $t$-Integrity} (\autoref{th:integrity}): For a DL task $\mathcal{F}$, input $I$, the output given by running the task locally is $O=\mathcal{F}(I)$. The outsourcing scheme with \name{} gives a output $O^*$. 
Then $Pr\left[O^* \notin \{O, \bot \}\right] < t$, where $\bot$ means the system aborts. 
\end{enumerate}

\section{\name{}}
\label{sec:details}

In this section, we first introduce the preliminary technique, then expound the concrete training procedure, and finally extend \name{} to distributed scenario. 
We concentrate on DNN training because inference is relatively simple and ``contained'' in training since it only involves the forward pass.

\noindent \textbf{Overview.} In \name{} (see~\autoref{fig:system}), the CSP comprises multiple nodes, including a master and several workers. The master is responsible for privacy protection and task assignments, and the workers execute computation tasks assigned by the master.
The master has two components, the \emph{task manager} (which is a TEE enclave), and an \emph{off-loader}. The task manager's TEE is the only trusted component in \name{}, where model and inputs are considered secure. All traffic involving the task manager is first encrypted inside the TEE (will be expound in~\autoref{subsec:enc}). The off-loader is designed to take charge of marshalling communications with the workers’ \emph{relay}s.
All inputs from the client are sent directly to the task manager inside the master via secure TLS channels.
To establish the secure channel, the task manager's TEE generates a TLS private key and binds the its code to the corresponding public key through remote attestation~\cite{johnson2016intel}. The client verifies the \emph{remote attestation} in a \emph{trust-on-first-use (TOFU)} security model.

\noindent \textbf{Key idea.}
According to previous literature~\cite{tramer2018slalom,agarap2018deep}, linear operations in DNNs, including fully-connected (FC) layers, convolutions, \etc, dominate most of the computational resources.
A linear operation performs a specific linear transformation from the input feature space to the output feature space, mathematically known as a linear map $L: \mathbb{R}^n \rightarrow \mathbb{R}^m$.
It is always possible to use \emph{matrix multiplications} to represent linear maps defined in finite-dimensional vector spaces~\cite{rudin1976principles}, \ie, $\exists \mathbf{A} \in \mathbb{R}^{m\times n}:L(\mathbf{x}) = \mathbf{A}\mathbf{x}$, where $\mathbf{A}$ is a real matrix.
Therefore, \textbf{all linear operations} in DNNs can be represented as matrix multiplications.
From now on, we consider all linear operations as matrix multiplications (denoted by $\times$ or omitted) without discrimination:
\begin{equation}
\label{eq:transformation}
\mathbf{Z}_l = \linear{\mathbf{W}_l}{\mathbf{X}_l} \sim \mathbf{W}_l \times \mathbf{X}_l. 
\end{equation}
\name{} offloads these linear operations to GPU-equipped workers for faster computation. We design an \emph{obfuscation algorithm} to blind both weights and input features before sending them to workers to preserve privacy.
The task manager only processes non-linear, and encryption/decryption operations during the training process.

\subsection{The \enc{} Algorithm}
\label{subsec:enc}

\begin{figure}
  \centering
  \includegraphics[width=0.55\linewidth]{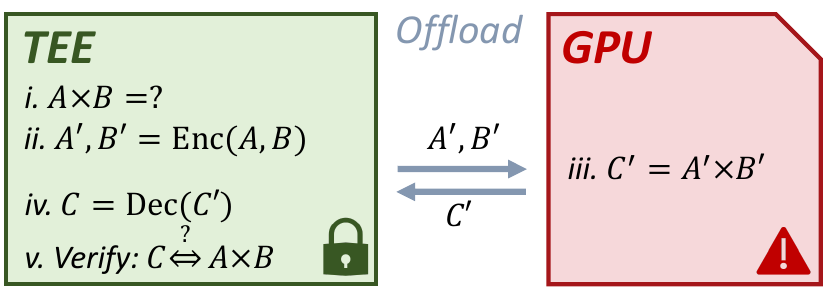}
  \vspace{-0.2cm}
  \caption{Workflow of the \enc{} algorithm in \name{}.}
  \label{fig:encryption}
\end{figure}

Offloading the computationally intensive matrix multiplications to GPUs facilitates fast batch computation. However, exposing the plaintext matrices outside the secure TEE induces risk of privacy breaches.
Hence, our objective is to \emph{obfuscate} the matrix operands before offloading them to GPU-equipped workers for computation.
We propose an \enc{} algorithm for \name{}, which is a form of permutation-based secure matrix multiplication \emph{optimized} for DNN tasks. 
The high-level workflow of this algorithm is illustrated in~\autoref{fig:encryption}.
The TEE offloads the computation $\mathbf{A} \times \mathbf{B}$.
The operands of the matrix multiplication $\mathbf{A},\mathbf{B}$ are first encrypted within the TEE (\autoref{alg:encryption}), and then the worker performs vanilla matrix multiplication directly on the encrypted matrices $\mathbf{A}',\mathbf{B}'$. The result $\mathbf{C}'$ is returned to the TEE in encrypted form and can be decrypted and verified to recover the plaintext $\mathbf{C}$ (\autoref{alg:decryption}).

\begin{algorithm}[hbt!]
\caption{$\texttt{KGen}(m,n,p)$---Secret Key Generation}
\fontsize{8}{9.5}\selectfont
\label{alg:skgen}
\DontPrintSemicolon
\SetNoFillComment
\KwData{Integers $m,n,p$}
\KwResult{A secret key $\mathsf{sk}$}
\tcc{Generate random coefficients}
$\mathbf{c_{m}} \overset{{\scriptscriptstyle\$}}{\gets} \{\mathbb{K}\}^m$,
$\mathbf{c_{n}} \overset{{\scriptscriptstyle\$}}{\gets} \{\mathbb{K}\}^n$,
$\mathbf{c_{p}} \overset{{\scriptscriptstyle\$}}{\gets} \{\mathbb{K}\}^p$\;
\tcc{Generate random permutations}
$\pi_m \gets \texttt{RandPr}(1,2,\cdots,m)$,
$\pi_n \gets \texttt{RandPr}(1,2,\cdots,n)$,
$\pi_p \gets \texttt{RandPr}(1,2,\cdots,p)$\;
\Return $\mathsf{sk}\gets(\underline{\mathbf{c_{m}}, \pi_m}, \underline{\mathbf{c_{n}},\pi_n}, \underline{\mathbf{c_{p}},\pi_p})$\tcc*[r]{$\mathsf{sk} \coloneqq (\mathsf{sk}_{\left[0\right]},\mathsf{sk}_{\left[1\right]},\mathsf{sk}_{\left[2\right]})$}
\end{algorithm}

\begin{algorithm}[hbt!]
\caption{$\texttt{Enc}(\mathsf{sk},\mathbf{A},\mathbf{B})$---Matrix Encryption}
\fontsize{8}{9.5}\selectfont
\label{alg:encryption}
\DontPrintSemicolon
\SetNoFillComment
\KwData{Two matrices, $\mathbf{A} \in \mathbb{R}^{m \times n}$ and $\mathbf{B} \in \mathbb{R}^{n \times p}$ and $\mathsf{sk}=\texttt{KGen}(m,n,p)$}
\KwResult{The encrypted matrices $\mathbf{A}', \mathbf{B}'$}
\For{$i \gets 1:m$ and $j \gets 1:n$}{
$\mathbf{A}(\cdots,i,j) \gets  \frac{\mathbf{c_{m}}(i)}{\mathbf{c_{n}}(j)}\mathbf{A}(\cdots,\pi_{m}(i),\pi_{n}(j)) $
}
\For{$i \gets 1:n$ and $j \gets 1:p$}{
$\mathbf{B}'(\cdots,i,j) \gets  \frac{\mathbf{c_{n}}(i)}{\mathbf{c_{p}}(j)}\mathbf{B}(\cdots,\pi_{n}(i),\pi_{p}(j)) $
}
\Return $(\mathbf{A}',\mathbf{B}')$
\end{algorithm}

We describe the encryption algorithm $\texttt{Enc}()$ in~\autoref{alg:encryption}. 
The matching decryption algorithm $\texttt{Dec}()$ is described in~\autoref{alg:decryption}, which uses the same set of secret key $\mathsf{sk}$ generated by~\autoref{alg:skgen} ($\texttt{KGen}()$) before encryption. In~\autoref{alg:skgen}, $\overset{{\scriptscriptstyle\$}}{\gets}$ is the random sampling function, $\mathbb{K} \in \mathbb{N}^+$ is the coefficient key space 
($\overset{{\scriptscriptstyle\$}}{\gets} \{\mathbb{K}\}^m$ means random sampling a vector with length $m$ from key space $\mathbb{K}$), and
$\texttt{RandPr}()$ is the random permutation generator~\cite{durstenfeld1964algorithm} ($\texttt{RandPr}(1,2,\cdots,n)$ returns a random permutation of the $n$ pre-images (mathematical term)).

\autoref{alg:decryption} describes the decryption and verification, in which $\pi^{-1}$ is the inverse permutation of $\pi$.
The probabilistic randomized Freivalds’ algorithm~\cite{freivalds2005fast} is adopted for result verification (\autoref{line:veri-start} to \autoref{line:veri-end}, see~\autoref{th:integrity} for setting of $k$).
We defer the proof of correctness of these algorithms to~\autoref{th:mm-correctness} and complexity analysis to~\autoref{subsec: cost-analysis}. 
The \enc{} algorithm is a vital technique to safeguard DL privacy and is continuously invoked during the training procedure.

\begin{algorithm}[hbt!]
\caption{$\texttt{Dec}(\mathsf{sk},\mathbf{C}')$---Matrix Decryption}
\fontsize{8}{9.5}\selectfont
\label{alg:decryption}
\DontPrintSemicolon
\SetNoFillComment
\KwData{The encrypted matrix $\mathbf{C}' \in \mathbb{R}^{m \times p}$, where $\mathbf{C}' = \mathbf{A}' \times \mathbf{B}'$ and $\mathsf{sk}=\texttt{KGen}(m,n,p)$}
\KwResult{The decrypted matrix $\mathbf{C}_{Dec}$ if it is correct}
\For(\tcc*[f]{Calculate decrypted matrix}){$i \gets 1:m$ and $j \gets 1:p$}{
${\mathbf{C}_{Dec}}(\cdots,i,j) \gets  \frac{\mathbf{c_{p}}({\pi_{p}^{-1}}(j))}{\mathbf{c_{m}}({\pi_{m}^{-1}}(i))}\mathbf{C}'(\cdots,{\pi_{m}^{-1}}(i),{\pi_{p}^{-1}}(j)) $
}
\For(\tcc*[f]{Integrity verification ($\mathbf{A},\mathbf{B}$ are kept inside the TEE from~\autoref{alg:encryption})}){$i \gets 1:k$}{\label{line:veri-start}
$\mathbf{r} \overset{{\scriptscriptstyle\$}}{\gets} \{0,1\}^p$\;
$\mathbf{P} \gets \mathbf{A}(\mathbf{B}\mathbf{r}^\top) - \mathbf{C}_{Dec}\mathbf{r}^\top$\tcc*[r]{Calculate $\mathbf{B}\mathbf{r}^\top$ first}
\If{$\mathbf{P} \neq \mathbf{0}_{m \times 1}$}{
Output ``failed'' and abort\;
}
}
Output ``passed'' and \Return $\mathbf{C}_{Dec}$\;
\label{line:veri-end}
\end{algorithm}

\subsection{Privacy-preserving Feedforward}
\label{subsec:feedforward}

In the next two sections, we first demonstrate the training method when there is only one worker, and we will discuss the case of multiple workers in~\autoref{subsec:dist}.

Noticing that in~\autoref{eq:feed-forward-batch}, there is one set of computational intensive linear operation (\ie, $\linear{\mathbf{W}_l}{\mathbf{X}_l}$) we would like to offload to GPU at layer $l$. 
The batched computation helps decrease the encryption/decryption cost as \name{} only requires one $\texttt{Enc}()$ at a layer for a mini-batch in the forward pass. 
The computations and communications of \name{} in the forward pass are shown in~\autoref{eq:feed-forward-name}.
Assumes that $\mathbf{W}_{l} \in \mathbb{R}^{m_l \times n_l}, \mathbf{X}_{l} \in \mathbb{R}^{n_l \times p_l}$\footnote{In PyTorch, tensors can have dimensions exceeding 2. In such cases, our algorithm consistently operates on the \textbf{last two dimensions}.}. 
Here, $\mathsf{sk}_{l}^{\text{Fwd}}$ is the secret key at layer $l$ generated in the forward pass using $\texttt{KGen}(m_l, n_l, p_l)$. In \name{}, a single set of $\mathsf{sk}_{l}^{\text{Fwd}}$ can be reused for all batches on this layer in an epoch once generated.
The task manager blind the two operands, $\mathbf{W}_l$ and $\mathbf{X}_l$, using $\texttt{Enc}()$ before offloading directly to GPU for matrix multiplication.
After obtaining $\mathbf{Z}_l$, TEE will compute $\mathbf{Y}_l$ (the non-linear operation) and pass it as the input $\mathbf{X}_{l+1}$ to the upcoming layer until the output layer $L$ is reached. The final output of the forward pass is $\mathbf{Y}_L$.

\NewDocumentCommand{\mysendmessageleft}{m}{%
  \sendmessage{<-,inner sep=0pt}{
    length=1cm,
    topstyle={inner sep=2pt},
    top=\smash[t]{$\scriptstyle#1$},
  }%
}
\NewDocumentCommand{\mysendmessageright}{m}{%
  \sendmessage{->,inner sep=0pt}{
  length=1cm,
    topstyle={inner sep=2pt},
    top=\smash[t]{$\scriptstyle#1$},
  }%
}
\newcommand{\smalleq}{\fontsize{7}{0}\selectfont}

\subsection{Privacy-preserving Backpropagation}
\label{subsec:backpropagation}

\begin{figure}
  \centering
  \includegraphics[width=0.9\linewidth]{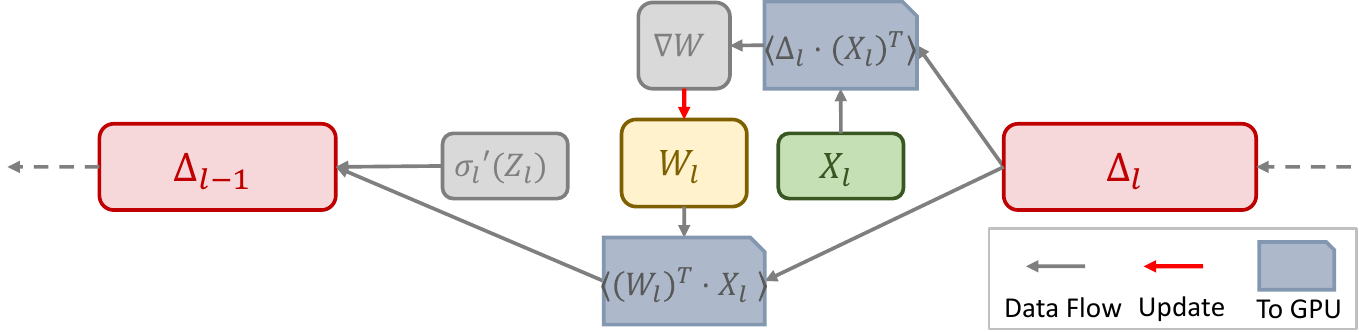}
  \vspace{-0.2cm}
  \caption{Backpropagation of DNN training in \name{}.}
  \label{fig:back}
\end{figure}

The computational graph of BP is depicted in~\autoref{fig:back}. At layer $l$, there are two sets of linear operations to be offloaded:
\begin{equation}
\label{eq:back-linearop} 
\begin{aligned}
    \linear{\Delta_{l}}{\mathbf{X}_{l}^\top}, \linear{\mathbf{W}_{l}^\top}{\Delta_{l}}.
\end{aligned}   
\end{equation}
Likewise, the simplest way is to generate two sets of $\mathsf{sk}_{l}$ and invoke~\autoref{alg:encryption} to obfuscate two sets of operands ($4$ matrices). 
However, the calculation of the encryption matrix incurs undesired overhead, and \name{} strives for minimizing the encryption/decryption cost as much as possible.

\noindent \textbf{Optimization of \enc{} in BP.}
Here, we observe two key insights that can help us optimize our encryption strategy:
\begin{enumerate}
    \item In~\autoref{eq:back-linearop}, it is only necessary to encrypt $3$ operand matrices instead of $4$ (\ie, when $\texttt{Enc}()$ is called directly twice), they are $\mathbf{W}_{l}^\top, \mathbf{X}_{l}^\top$ and $\Delta_{l}$.
    \item Recall in the forward pass, we have used $\mathsf{sk}_{l}^{\text{Fwd}}$ to calculate the encrypted matrices $\mathbf{W}_{l}'$ and $\mathbf{X}_{l}'$ already. 
\end{enumerate}

\begin{figure}
  \centering
  \includegraphics[width=0.85\linewidth]{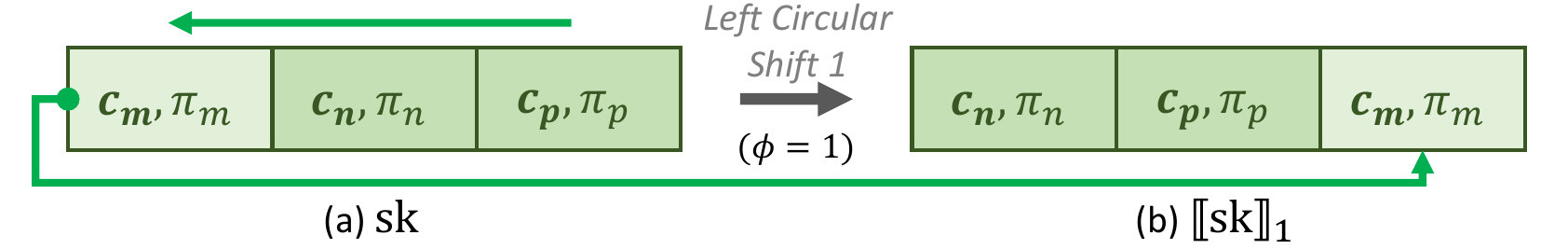}
  \vspace{-0.2cm}
  \caption{An example of left circular key shift used in backpropagation.}
  \label{fig:keyshift}
\end{figure}

A natural question is: \emph{Can we reuse the encrypted results instead of recomputing the encryption?}
Fortunately, it is possible to reuse the encrypted matrices from the forward pass. 
Recall that in~\autoref{alg:skgen}, the secret key $\mathsf{sk}_{l}^{\text{Fwd}}$ used for encryption depends on the shape of the matrix operands.
We observe that $\Delta_{l} \in \mathbb{R}^{m_l \times p_l}$ has the same shape with $\mathbf{Z}_l$.
We now define a left circular key shift $\keyshift{\mathsf{sk}}{\phi}$ where $\mathsf{sk}=\texttt{KGen}(m,n,p)$. 
First, we divide the components of $\mathsf{sk}$ into three ordered tuples, $\mathsf{sk}_{\left[0\right]}=(\mathbf{c_m},\pi_m), \mathsf{sk}_{\left[1\right]}=(\mathbf{c_n},\pi_n), \mathsf{sk}_{\left[2\right]}=(\mathbf{c_p},\pi_p)$. 
Then, the left circular key shift is defined as $(\keyshift{\mathsf{sk}}{\phi})_{\left[i\right]} \triangleq \mathsf{sk}_{\left[(i-\phi)\mod3\right]}$. 
A demonstration of the key shift with $\phi = 1$ is exemplified in~\autoref{fig:keyshift}.
Now, it is possible to use $\keyshift{\mathsf{sk}_{l}^{\text{Fwd}}}{2}$ to compute the encrypted version of $\Delta_{l}^\top$ using $\texttt{Enc}(\keyshift{\mathsf{sk}_{l}^{\text{Fwd}}}{2}, \Delta_{l}^\top, \cdot)$ without passing the second matrix as parameter (``$\cdot$'' denotes an \emph{omitted} parameter), and reusing $\mathbf{W}_{l}', \mathbf{X}_{l}'$ in the offloaded matrix multiplications.
The computations and communications of \name{} in BP are shown in~\autoref{eq:back-name}.
Here, $\mathbf{Z}_{l}, \mathbf{X}_{l}', \mathbf{W}_{l}'$ are reused from the forward pass, in which $\mathbf{Z}_{l}$ is kept in the secure memory of the TEE and $\mathbf{X}_{l}', \mathbf{W}_{l}'$ are stored in the worker's GPU memory.
As with the forward pass, \name{} requires only one $\texttt{Enc}()$ per layer in BP. 
Also, the two $\texttt{Dec}()$s are already the minimum decryption cost, since there are two sets of linear operations.

\setlength{\abovedisplayskip}{0.25cm}
\begin{figure}
    \centering
\pseudocodeblock[linenumbering,skipfirstln]{ 
 \textbf{Task Manager (TEE)}  \<\< \textbf{Worker (GPU)} \\[0.1\baselineskip][\hline]
 \rule{0pt}{1\normalbaselineskip}
\smalleq{
 (\textbf{W}_l', \textbf{X}_l') = \texttt{Enc}(\textsf{sk}_{l}^{\text{Fwd}}, \textbf{W}_l, \textbf{X}_l),}\\
 \<\mysendmessageright{\textbf{W}_l', \textbf{X}_l'} \<\smalleq{\textbf{Z}_l' = \textbf{W}_l'\textbf{X}_l',}\\
  \smalleq{\textbf{Z}_l = \texttt{Dec}(\textsf{sk}_{l}^{\text{Fwd}}, \textbf{Z}_l'),}\<\mysendmessageleft{\textbf{Z}_l'} \< \smalleq{(\text{Keep } \textbf{W}_l'\textbf{X}_l' \text{ in mem.})}\\
  \smalleq{\textbf{X}_{l+1} = \sigma_l(\textbf{Z}_l).}\pclb[0.2\baselineskip]\hline
}
    \caption{Computations of \name{} in forward pass at layer $l$}
    \label{eq:feed-forward-name}
\end{figure}

\begin{figure}
    \centering
\pseudocode[linenumbering,skipfirstln,bodylinesep=-5pt]{
 \textbf{Task Manager (TEE)}  \<\< \textbf{Worker (GPU)} \\[0.1\baselineskip][\hline]
\rule{0pt}{1\normalbaselineskip}
\smalleq{
 (\Delta_l^\top)'= \texttt{Enc}(\keyshift{\textsf{sk}_{l}^{\text{Fwd}}}{2}, \Delta_{l}^\top, \cdot),}\\
 \<\mysendmessageright{(\Delta_l^\top)'} 
 \<\smalleq{
    (\textbf{T}_l^1)' = \textbf{X}_l'(\Delta_l^\top)',}\\
 \<\<\smalleq{
    (\textbf{T}_l^2)' = (\Delta_l^\top)'\textbf{W}_l',}\\
  \smalleq{\textbf{T}_l^1 = \texttt{Dec}(\keyshift{\textsf{sk}_{l}^{\text{Fwd}}}{1}, (\textbf{T}_l^1)'),} \<\mysendmessageleft{(\textbf{T}_l^1)',(\textbf{T}_l^2)'} \\
  \smalleq{\textbf{T}_l^2 = \texttt{Dec}(\keyshift{\textsf{sk}_{l}^{\text{Fwd}}}{2}, (\textbf{T}_l^2)'),} \\
  \smalleq{\textbf{W}_l^* \gets \textbf{W}_l - \frac{\lambda}{\mathcal{B}}(\textbf{T}_l^1)^\top,}\\
  \smalleq{\Delta_{l-1}^\top = \textbf{T}_l^2\odot\sigma'_{l}(\textbf{Z}_{l}^\top).}\pclb[0.2\baselineskip]\hline
}
    \caption{Computations of \name{} in backpropagation at layer $l$}
    \label{eq:back-name}
\end{figure}

\subsection{Distributed Training using \name{}}
\label{subsec:dist}

\begin{figure}
  \centering
  \includegraphics[width=\linewidth]{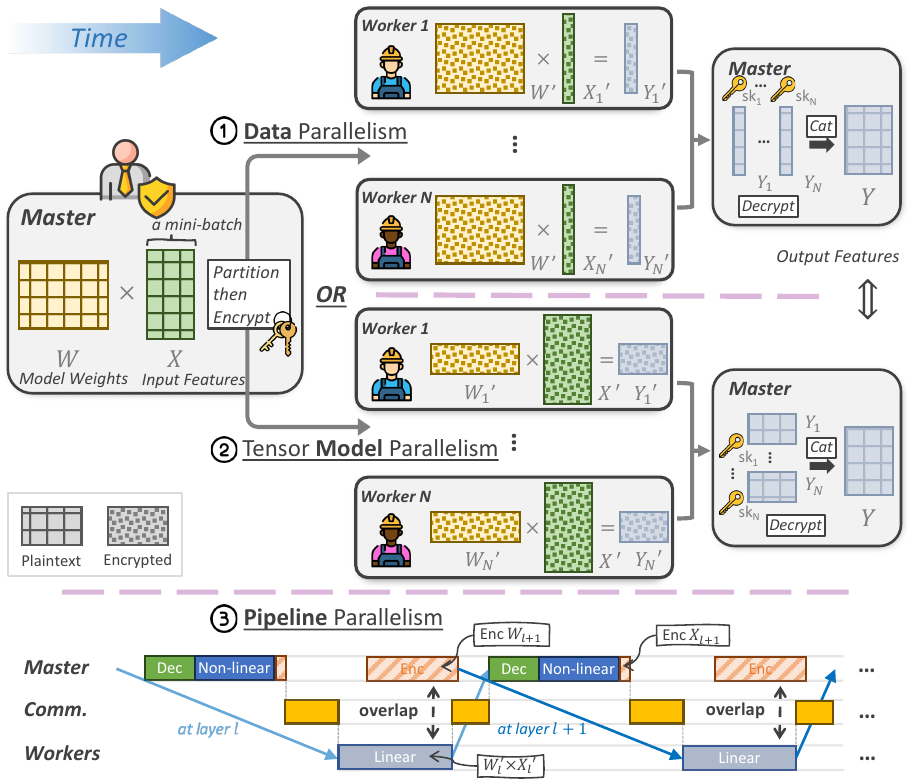}
  \vspace{-0.5cm}
  \caption{Distributed parallelism used in \name{}.}
  \label{fig:parallelism}
\end{figure} 

Now we elaborate the training strategy using \name{} when there are multiple workers.
This distributed training method can be scaled to larger models in the future via \textbf{hybrid distributed parallelism} combining DP, TP, and PP (illustrated in \autoref{fig:parallelism}).
We specify the complete training strategy in~\autoref{alg:training}. 

\emph{Data parallelism (DP):} The matrix multiplication is replicated to each worker with each being fed a slice of the partitioned data. 
When all the calculations are complete, a \texttt{Gather} operation is triggered and the results are \emph{aggregated, decrypted and concatenated} inside the master. As the weights need to be copied to all workers, DP incurs substantial communication overhead as the model size gets larger.
\emph{Tensor model parallelism (TP):} Model weights are partitioned across multiple workers with each receiving a part of the model and conducting part of the computation. 
\emph{Pipeline parallelism\footnote{This is different from the commonly-known microbatch pipeline~\cite{narayanan2019pipedream}.} (PP):} To minimize the encryption cost, we further adopt \emph{pipelined encryption} and asynchronous SGD update, which enables the master to encrypt weights at the upcoming layer in waiting for the results handled by the workers. 
In this way, we further alleviate the encryption overhead by overlapping it with communication and computation.

For efficient training on distributed workers, \name{} always use PP, and switch DP and TP alternately \emph{depending on the properties of the linear operation}:
(a) Fully-connected (FC) layers are parameter-heavy, use TP to partition the model weights;
(b) Convolutional layers have fewer parameters but is computed on larger input features, use DP to distribute the data;
(c) For the unique linear operation of the Transformer~\cite{vaswani2017attention}, multi-head attention (MHA), it is natural to use TP because multiple independent heads can be segmented.
We point out that \name{} can collaborate with other distributed training algorithms in an \emph{ad-hoc} way to further boost performance, which can be an interesting direction for future work.

\begin{algorithm}[hbt!]
\caption{Distributed DNN Training in \name{}}
\fontsize{8}{9.5}\selectfont
\label{alg:training}
\DontPrintSemicolon
\SetNoFillComment
\KwData{Training dataset $\mathbf{D} = (\bar{\mathbf{X}},\bar{\mathbf{Y}})$, 
DNN framework $(\mathbf{W}, \sigma, L)$, 
loss function $\mathcal{J}$, 
number of workers $\mathcal{N}$
}
\KwResult{The trained model $\mathbf{W}^*$}

\For{all epoch}{
\For(\tcc*[f]{$|\mathbf{X}_1|=batch\_size$}){all mini-batch $(\mathbf{X}_1,\mathbf{Y}) \in \mathbf{D}$}{
\For(\tcc*[f]{Forward pass (inference)}){$l \gets 1:L$}{
\label{line:start}
Partition weights $\mathbf{W}_l$ \emph{or} input $\mathbf{X}_l$ to $\mathcal{N}$ shares depending on the type of the layer, and get \emph{partitioned} operands $\mathbf{W}_{l(,j)}$ and $\mathbf{X}_{l(,j)}$, $j=1,\cdots,\mathcal{N}$.\;
Generate/refresh secret keys $\mathsf{sk}_{l,j}$ using~\autoref{alg:skgen}.\;
Encrypt $\mathcal{N}$ shares of operands, distribute them to the workers following~\autoref{eq:feed-forward-name} to obtain $\mathbf{X}_{l+1}$.\;
The workers stores the encrypted operands $(\mathbf{W}_{l(,j)})'$ and $(\mathbf{X}_{l(,j)})'$ in memory.\tcc*[r]{Keep encrypted results for reuse in BP}
}
\label{line:end}
$\Delta_{L} \gets \nabla\mathcal{J}(\mathbf{X}_{L+1}, \mathbf{Y}) \odot \sigma'_{l}(\mathbf{Z}_{L})$\;
\For(\tcc*[f]{Backpropagation (BP)}){$l \gets L:1$}{
Partition $\Delta_l$ \textbf{if} the corresponding input $\mathbf{X}_l$ is partitioned in the forward pass.\;
Encrypt $\mathcal{N}$ shares of operands, distribute them to the workers following~\autoref{eq:back-name} to update $\mathbf{W}_l^*$.\;
}
}
}
\Return $\mathbf{W}^*$
\end{algorithm}

\section{Analysis of \name{}}
\label{sec:analysis}

\subsection{Proof of Properties of \name{}}

\begin{theorem}[Correctness]
\label{th:mm-correctness}
The \enc{} algorithm (described in \autoref{alg:encryption} and \autoref{alg:decryption}) is correct, i.e., let $(\mathbf{A}',\mathbf{B}') = \texttt{Enc}(\mathsf{sk},\mathbf{A},\mathbf{B})$ and $\mathbf{C}' = \mathbf{A}'\mathbf{B}'$. Then, $\texttt{Dec}(\mathsf{sk}, \mathbf{C}') = \mathbf{A}\mathbf{B}$.
\end{theorem}
\begin{proof}
We start with constructing \emph{encryption matrices} using $\mathsf{sk}=(\mathbf{c_{m}}, \pi_m, \mathbf{c_{n}},\pi_n, \mathbf{c_{p}},\pi_p)$ generated from~\autoref{alg:skgen}:
\[
\left\{
\begin{aligned}
&\mathbf{E}_1(i,j) = \mathbf{c_{m}}(i)\delta_{\pi_m(i),j},\\
&\mathbf{E}_2(i,j) = \mathbf{c_{n}}(i)\delta_{\pi_n(i),j},\\
&\mathbf{E}_3(i,j) = \mathbf{c_{p}}(i)\delta_{\pi_p(i),j},
\end{aligned}
\right.
\] 
in which $\scriptsize
\delta_{i,j}=\left\{
\begin{aligned}
1, &  & i = j \\
0, &  & i \neq j
\end{aligned}
\right.$ is the Kronecker delta function.
Note that, the encryption matrices are invertible. This is due to $0 \notin \mathbf{c_{m}}$, thus the determinant $det(\mathbf{E}_1) \neq 0$. Similar proof also applies to $\mathbf{E}_2, \mathbf{E}_3$. 
The inverse encryption matrices are:
\[
\left\{
\begin{aligned}
&{\mathbf{E}_1^{-1}}(i,j) = {\mathbf{c_{m}}(i)}^{-1}\delta_{\pi^{-1}_m(i),j},\\
&{\mathbf{E}_2^{-1}}(i,j) = {\mathbf{c_{n}}(i)}^{-1}\delta_{\pi^{-1}_n(i),j},\\
&{\mathbf{E}_3^{-1}}(i,j) = {\mathbf{c_{p}}(i)}^{-1}\delta_{\pi^{-1}_p(i),j}.
\end{aligned}
\right.
\]
Recall that in~\autoref{alg:encryption}, the encrypted matrix ${\mathbf{A}'}(i,j) =  \frac{\mathbf{c_{m}}(i)}{\mathbf{c_{n}}(j)}\mathbf{A}(\pi_{m}(i),\pi_{n}(j))$. Hence, the encrypted matrix can be represented as $\mathbf{A}' = \mathbf{E}_1\mathbf{A}\mathbf{E}_2^{-1}$. 
Likewise, $\mathbf{B}' = \mathbf{E}_2\mathbf{B}\mathbf{E}_3^{-1}$.
Therefore, for the encrypted matrix multiplication:
\begin{equation}
\label{eq:proof}
\begin{aligned}
\mathbf{A}'\mathbf{B}' &= (\mathbf{E}_1\mathbf{A}\mathbf{E}_2^{-1})(\mathbf{E}_2\mathbf{B}\mathbf{E}_3^{-1}) \\&= \mathbf{E}_1\mathbf{A}\mathbf{B}\mathbf{E}_3^{-1} = \mathbf{C}'.
\end{aligned}
\end{equation}
By ~\autoref{alg:decryption}, the decrypted matrix can be represented using the aforementioned encryption matrices as $\mathbf{C}_{Dec}=\texttt{Dec}(\mathsf{sk}, \mathbf{C}') = \mathbf{E}_1^{-1}\mathbf{C}'\mathbf{E}_3$. Due to~\autoref{eq:proof}, the following equation holds:
\begin{equation*}
\mathbf{C}_{Dec} = \mathbf{E}_1^{-1}\mathbf{C}'\mathbf{E}_3=\mathbf{E}_1^{-1}(\mathbf{E}_1\mathbf{A}\mathbf{B}\mathbf{E}_3^{-1})\mathbf{E}_3=\mathbf{A}\mathbf{B}.
\qedhere
\end{equation*}
\end{proof}

\begin{theorem}[Integrity] 
\label{th:integrity}
\name{} achieves verifiable $t$-integrity (see~\autoref{subsec: goals}), by setting $k >  \log_2{\frac{1}{1-(1-t)^{\frac{1}{\alpha \mathcal{N}L}}}}$ in~\autoref{alg:decryption} where $\scriptsize\alpha=\left\{
\begin{aligned}
&1, &  \mathcal{F}=\text{inference} \\
&3n{\tiny\lceil\frac{|\mathbf{D}|}{\mathcal{B}}\rceil}, &  \mathcal{F}=\text{training}
\end{aligned}
\right.$, and $n$ is the number of training epochs.
\end{theorem}
\begin{proof}
Since the untrusted components can be adversarial, \name{} verifies the integrity of offloaded results using the probabilistic Freivalds’ algorithm~\cite{freivalds2005fast}, which produces a soundness one-sided error (no false negatives) of $\frac{1}{2^k}$, \ie, {\small $P_{\text{err}}=Pr\left[\mathbf{C}_{Dec} \neq \mathbf{A}\mathbf{B} | \autoref{alg:decryption} \text{ Outputs ``passed''}\right] \le \frac{1}{2^k}$}.
Setting {\small$k > \log_2{\frac{1}{1-(1-t)^{\frac{1}{\alpha \mathcal{N}L}}}}$}, we have {\small $P_{\text{err}} < 1-(1-t)^{\frac{1}{\alpha \mathcal{N}L}}$}. 
According to algorithms in ~\autoref{eq:feed-forward-name} and~\autoref{eq:back-name}, there are $L$ and $2L$ offloaded computations for a batch and a single worker in an epoch in the forward pass and backpropagation, respectively. 
The number of offloaded computations in $\mathcal{F}$ is $\alpha \mathcal{N}L$.
Thus we have that:

\begin{small}
\begin{equation}
\begin{aligned}
    Pr\left[O^* \neq \mathcal{F}(I) \vert O^* \neq \bot \right] &\leq 1 - (1-P_{\text{err}})^{\alpha \mathcal{N}L}\\
    &< 1- (1-t) = t.
\end{aligned}
\end{equation}
\end{small}
Thus the following inequality holds:

\begin{small} 
\begin{equation*}
\begin{aligned}
    Pr\left[O^* \notin \{ \mathcal{F}(I),\bot \}\right] &= Pr\left[O^* \neq \mathcal{F}(I) \cap O^* \neq \bot \right]\\
    &= Pr\left[O^* \neq \bot \right]Pr\left[O^* \neq \mathcal{F}(I) \vert O^* \neq \bot \right]\\  
    &\leq Pr\left[O^* \neq \mathcal{F}(I) \vert O^* \neq \bot \right] < t.
    \qquad \quad \qedhere 
\end{aligned}
\end{equation*}
\end{small}
\end{proof}

\begin{theorem}[Privacy]
\label{th:privacy}
    \name{} achieves privacy w.r.t. model and input (see~\autoref{subsec: goals}) with the proposed \enc{} algorithm.
\end{theorem}
\begin{proof}
    For \autoref{alg:encryption}, assuming that the generated permutations $\pi_m, \pi_n$ are truly random, then the complexity to reconstruct $\mathbf{A}$ from the encrypted matrices $\mathbf{A}' \in \mathbb{R}^{m \times n}$ is $\mathcal{O}(m!n!\left|\mathbb{K}\right|^{(m+n)})$, which means the expected time of an adversary launching a brute-force attack is definitely \emph{non-polynomial}.
    Thus, privacy \wrt{} model and input is achieved by encryption before offloading to untrusted components.
\end{proof}

\noindent\textbf{Remark.} \autoref{th:mm-correctness} suggests the correctness of our proposed \enc{} algorithm. \autoref{th:integrity} and \autoref{th:privacy} prove that \name{} achieves verifiable $t$-integrity and privacy \wrt{} model and input by integrating the algorithm.

\subsection{Measuring Privacy Level}

\begin{figure}
  \centering
  \includegraphics[width=\linewidth]{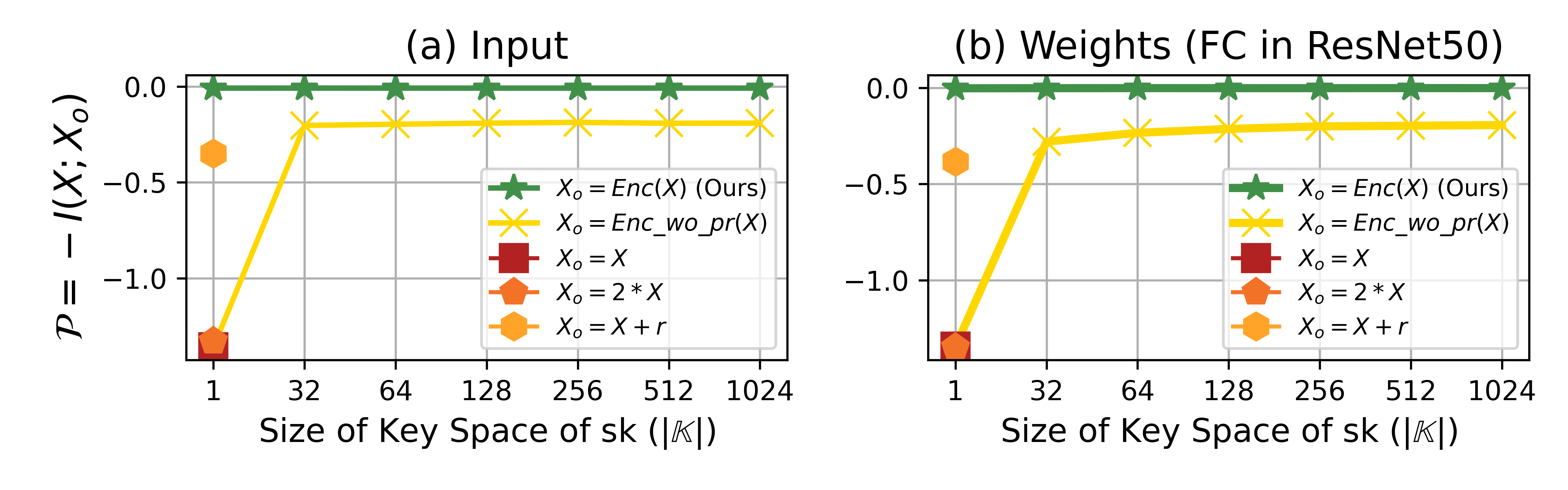}
  \vspace{-0.7cm}
  \caption{Privacy level $\mathcal{P}(\uparrow)$ of \name{} of (a) input and (b) weights. \name{}'s $\texttt{Enc}()$ achieves similar privacy close to $0$ \wrt{} different size of key space.}
  \label{fig:exp-mi}
\end{figure}

Mutual information (MI)~\cite{shannon2001mathematical} is a quantification of how similar or different two variables are. Let $(X, Y)$ be a pair of random variables with a joint distribution $p(X,Y)$ and their marginal distributions $p(X),p(Y)$,  MI is defined as:
\begin{equation}
\begin{aligned}
    I(X;Y) &= D_{KL}(p(X,Y) \Vert p(X)p(Y))\\
           &= H(X)-H(X \vert Y).
\end{aligned}    
\end{equation}
Where $D_{KL}$ is the Kullback–Leibler divergence, $H(X)$ and $H(X \vert Y)$ denote the marginal entropy and conditional entropy.
The MI $I(X;Y)$ is non-negative and symmetric. Intuitively, a lower MI indicates less amounts of common knowledge shared between the two variables $X$ and $Y$.
To quantify the privacy-preserving level of \name{}, we define the privacy as:
\begin{equation}
    \mathcal{P} = -I(X;X_o) = -I(X; \texttt{Enc}(\mathsf{sk},X,\cdot)).
\end{equation}
Here, $X$ is the \emph{private data} (\ie, weight $\mathbf{W}$ and input $\mathbf{X}_1$) and $X_o$ is the \emph{observable data} in the view of the adversary, which in terms of \name{} is the encrypted variable $\texttt{Enc}(\mathsf{sk},X,\cdot)$. 
In information-theoretic privacy, \emph{perfect privacy} is achieved if $\mathcal{P} = 0$.
To estimate the privacy level of \name{}, we depict the MI $I(X;X_o)$ between the encrypted data obtained by the proposed \enc{} algorithm along with other comparable schemes and the original input selected from CIFAR-10 (see~\autoref{sec:evaluation}) and model weights in~\autoref{fig:exp-mi}. 
Here, the x-axis is the size of key space $\left|\mathbb{K}\right|$, $\texttt{Enc}()$ is our proposed encryption algorithm, $\texttt{Enc\_wo\_pr}()$ is our algorithm without using permutations. We also show the MI of direct exposure, multiplication by scalar, and addition with random matrix ($X_o= X+r$, which is similar to techniques in~\cite{tramer2018slalom, hou2021model}).

\subsection{Encryption-decryption Cost Analysis}
\label{subsec: cost-analysis}
The encryption~\autoref{alg:encryption} yields a (time) complexity of $\mathcal{O}(mn+np)$. 
\autoref{alg:skgen} also consumes a negligible cost of $\mathcal{O}(m+n+p)$ for generating secret key $\mathsf{sk}_l$ at layer $l$, which can be reused over all the iterations.
The decryption~\autoref{alg:decryption} yields a total complexity of $\mathcal{O}(mp)$.
The \enc{} algorithm is meaningful since the matrix multiplication exhibits a complexity of $\mathcal{O}(mnp) \gg \mathcal{O}(mn+np+mp)$ when $m,n,p$ is large. 
TEE significantly circumvents the computationally heavy linear operations by offloading it to GPUs.

In addition, by reusing the encrypted values, the encryption complexity for training has reduced to $50\%$ roughly (forward: $2$ matrices to encrypt; BP: $1$ matrix instead of $4$ to encrypt) in one training epoch.
Since the cost of the \enc{} is proportional to the size of the matrix and the dimension of DNN parameters are substantially larger than those of the input, thus encrypting $\mathbf{W}_l$ (see~\autoref{eq:feed-forward-name}) sand decrypting $(\mathbf{T}_l^1)'$ (see~\autoref{eq:back-name}) are the most significant overhead.
However, this is a necessity because (a) we need to protect \emph{all} the weights; and (b) the complexity of our algorithm is proportional to the number of weights, which is already \emph{optimal} analytically.

\section{System Implementation}
\label{sec:implementation}

We implement \name{} with Intel SGX CPU as the TEE and NVIDIA GPUs as accelerators. 
We use Graphene-SGX~\cite{tsai2017graphene} (now renamed to Gramine\footnote{\url{https://github.com/gramineproject/gramine}}), a lightweight library OS designed for running an unmodified confidential application inside, to safeguard the task manager within an SGX enclave.
For the software stack, we choose PyTorch and its CPP extension\footnote{\url{https://pytorch.org/tutorials/advanced/cpp_extension.html}} for DNN and tensor-related implementation.
We implement our prototypes primarily using C++ and Python in $3,907$ LoC.

\name{}'s master handles the input provided by the client before initiating the DL task. 
This is implemented using a standard SGX remote attestation. Simply put, the client submits the model in \texttt{ONNX} format~\cite{lin2019onnc} and data to the master's TEE directly via a secure TLS channel. The model is then decrypted inside the TEE using standard TLS cipher suites.

We implement two prototypes of \name{} for different types of communication.
\emph{\name{}-l} (\underline{l}ocal) applies to the master and workers located on the same (physical) server.
\emph{\name{}-n} (\underline{n}etwork) applies to each nodes located on distributed servers in a LAN.
In both prototypes, the task manager spawns an extra PyTorch process to first offload the encrypted data to the off-loader via shared memory, this is due to the inconvenience of modifying the the syscalls inside the TEE~\cite{shen2022soter}. 
For \emph{\name{}-n}, the off-loader streams the data via TCP to the workers' relays. 
The relay then moves the data to GPU memory using \texttt{cudaMemcpy()}. We simulate network connection in \emph{\name{}-n} by packing the master and workers into \emph{separate} Docker containers, thus the nodes are inter-connected via virtual \texttt{Bridge}\footnote{\url{https://docs.docker.com/network/drivers/bridge/}} network. 
We use a Linux tool \texttt{qperf}\footnote{\url{https://github.com/linux-rdma/qperf}} to measure the network performance, the average (tested 10 times) TCP latency using bridge network between two containers is \SI{24.4}{\micro\second}.

\section{Performance Evaluation}
\label{sec:evaluation}

We evaluate our implementation on a physical server with Intel(R) Xeon(R) Gold 5218R CPU processor and \SI{376}{GB} of available RAM. This machine is co-allocated with $2$ NVIDIA GeForce RTX 4090 GPUs equipped with \SI{24}{GB} of VRAM each. 
We use the simulation mode of SGX for evaluation.

\noindent \textbf{Models, dataset, and baselines.}  
We select two DNNs, ResNet50~\cite{agarap2018deep}, a CNN with residual connections, and Vision Transformer (ViT)~\cite{dosovitskiy2020image}, a Transformer-based~\cite{vaswani2017attention} DNN. We conduct image classification, a classic supervised DL task, using the CIFAR-10 dataset\footnote{\url{https://www.cs.toronto.edu/~kriz/cifar.html}} which comprises $60,000$ images that span $10$ categories.
Since There is \emph{no} TEE-assisted system that guarantees both efficient DNN training (leveraging GPUs in BP) and model privacy protection, we compare our prototypes with three baseline systems: two \emph{secure baselines} that preserve both model and input privacy, including Pure TEE and MLCapsule~\cite{hanzlik2021mlcapsule} (inference only), they both shield an entire model within the TEE without employing GPUs; and DarKnight~\cite{hashemi2021darknight}, an \emph{insecure baseline} that does not protect model privacy.
We aim to answer the following questions:
\newcommand{\req}[1]{\emph{Q#1}}
\begin{enumerate}
    \item[\textbf{\req{1}.}] (\emph{Efficiency}) What is the training/inference performance of \name{}? 
    \item[\textbf{\req{2}.}] (\emph{Accuracy}) Does \name{} achieve the same level of accuracy as the baselines?
    \item[\textbf{\req{3}.}] (\emph{Empirical Privacy}) Would \name{} leak any sort of information when encounters real-world attacks?
\end{enumerate}

\begin{figure}
  \centering
  \includegraphics[width=0.95\linewidth]{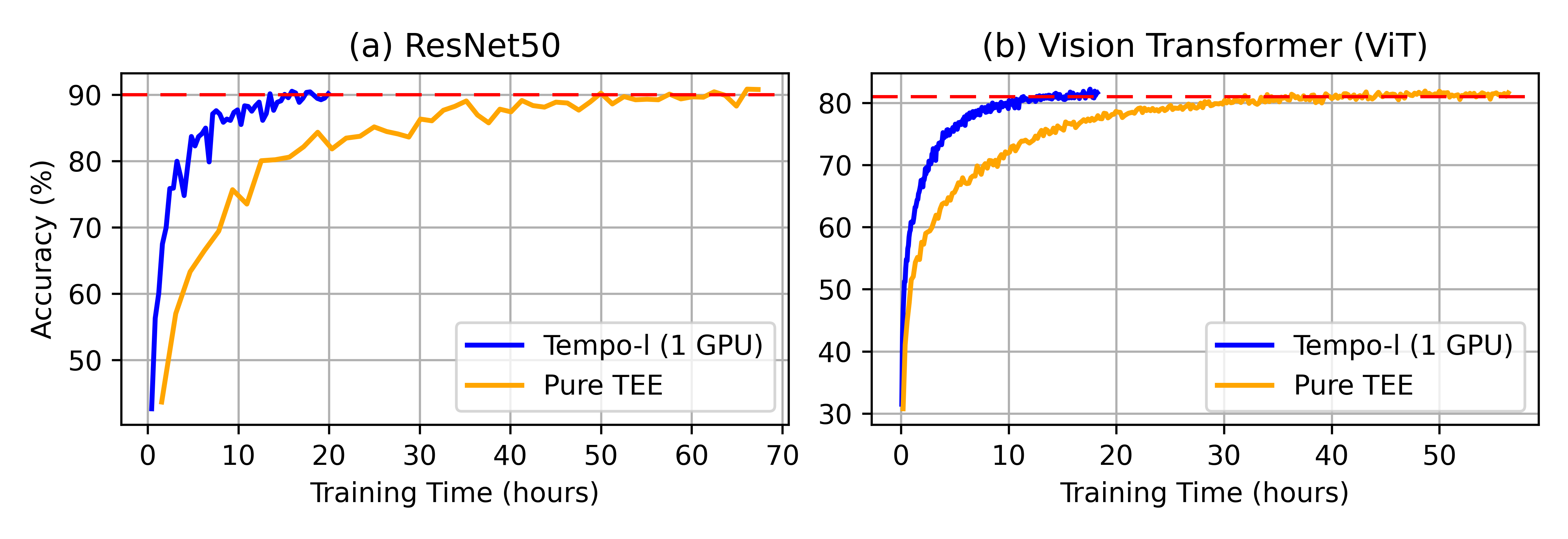}
  \vspace{-0.3cm}
  \caption{Training accuracy (on test set) of \emph{\name{}-l} (1 GPU) over CIFAR-10 for two DNNs: (a) ResNet50 (b) ViT, compared with the secure baseline.}
  \label{fig:exp-trian}
\end{figure} 

\begin{figure}
  \centering
  \includegraphics[width=\linewidth]{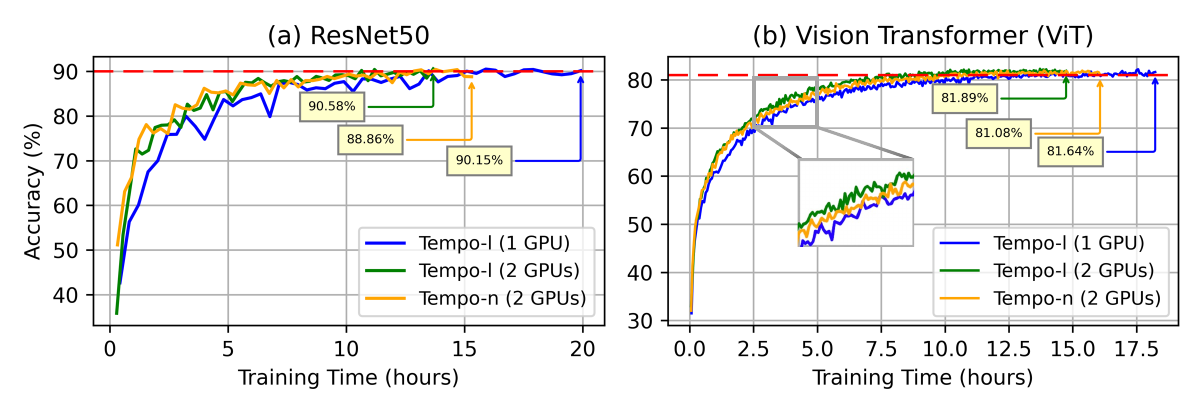}
  \vspace{-0.6cm}
  \caption{Distributed training accuracy (on test set) of \name{} over CIFAR-10 for two DNNs: (a) ResNet50 (b) ViT. \emph{\name{}-n} suggests that the master (TEE) communicates with the workers (GPU) using network.}
  \label{fig:exp-dst}
\end{figure}

\subsection{Training Performance}

\begin{figure}
  \centering
  \includegraphics[width=0.9\linewidth]{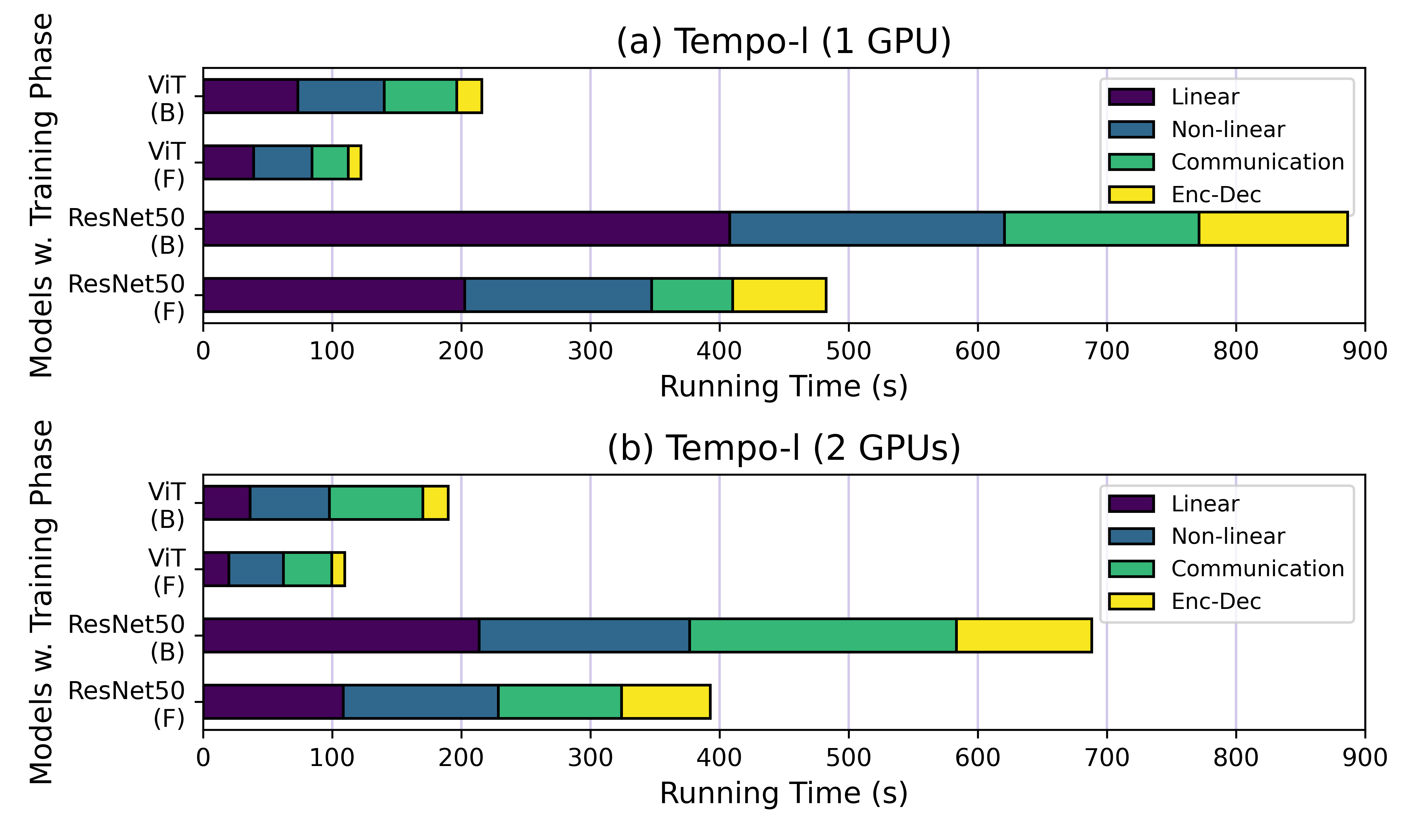}
  \vspace{-0.2cm}
  \caption{Training time breakdown. We measure the time cost averaged on an epoch. (F) stands for forward pass and (B) stands for backpropagation. We evaluate for two settings: \emph{\name{}-l} with (a) a single GPU (b) 2 GPUs.}
  \label{fig:exp-micro}
\end{figure}

For DNN training tests, we choose Pure TEE as the secure baseline and DarKnight~\cite{hashemi2021darknight} as the insecure baseline.
We trained two models over CIFAR-10, with $50,000$ images in the training set and the rest $10,000$ in the test set. 
We use mini-batch training with a learning rate of $0.001$ and the batch size set to $512$ images.
(\textbf{\req{2}}) The ResNet50 is trained for $50$ epochs to achieve a final accuracy of over $90\%$ on the test set while the ViT (patch size $=4$ and encoder layers $=8$, see~\cite{dosovitskiy2020image} for meanings) is trained for $300$ epochs to achieve a final accuracy of over $81\%$\footnote{The difference in accuracy of the two models may be due to the inherited characteristic of the models themselves. According to~\cite{dosovitskiy2020image}, ViT may underperform with less training data and without pre-training.}. 
The training curves of accuracy are depicted in~\autoref{fig:exp-trian}.
Please note that, within the same number of training epochs, \name{} achieves \emph{intact accuracy} when compared to the baseline. 
We also conduct experiments for distributed training over multiple workers (see~\autoref{fig:exp-dst}). For training using $2$ workers, we scale up the batch size to $1024$. Using multiple workers increases efficiency due to hybrid parallelism among GPUs, allowing a larger batch size thus reducing the number of iterations. 
However, the speed-up is nonlinear, this is due to the fact that the training procedure constitutes other operations in addition to linear operations. 
\autoref{fig:exp-micro} displays a detailed time breakdown of various operations. By adding workers, we observe that the time spent in linear operations is greatly reduced.
For \emph{\name{}-l} and \emph{\name{}-n} with the same number of workers, the latter incurs a noticeable slowdown due to the network latency among the nodes.

(\textbf{\req{1}} - Part I) We report the training time and corresponding speed-up of \name{} with baselines in~\autoref{tab:speedup}. 
\name{} yields a maximal training speed-up of $4.92 \times$ and $3.83 \times$ compared with the secure baseline for ResNet50 and ViT, respectively.
Out of curiosity, we also compare it with an insecure baseline, \ie, training is conducted with unprotected GPU without \emph{any} privacy protection using PyTorch directly.
No surprising that \name{} is an order of magnitude slower than the insecure baseline thanks to the engagement of TEE and communications. Please note that similar privacy-preserving training systems exhibit similar slowdowns even without the model protected~\cite{hashemi2021darknight,ng2021goten}.

\begin{table}
\centering\fontsize{7.75}{10}\selectfont
\setlength\tabcolsep{0.05cm}
\caption{Training Time (in hours) and Speed-up of \name{} over CIFAR-10, compared with baselines.}
\vspace{-0.15cm}
\begin{threeparttable}
\begin{tabularx}{\linewidth}{c*7{>{\centering\arraybackslash}X}}
\toprule
\textbf{Model} &\emph{\name{}-l} (1 GPU) &\emph{\name{}-l} (2 GPUs) &\emph{\name{}-n} (2 GPUs) &Pure TEE\tnote{1} &DarKnight\cite{hashemi2021darknight}\tnote{2} &Non-private\tnote{2}\\
\midrule
ResNet50 
&\makecell{19.91\\\textcolor{mygreen}{$3.38 \times$}} 
&\makecell{13.68\\\textcolor{mygreen}{$\textbf{4.92} \times$}}
&\makecell{15.30\\\textcolor{mygreen}{$4.40 \times$}}
&\makecell{67.32\\\textcolor{mygreen}{$1.00 \times$}} 
&\makecell{11.87\\\textcolor{myred}{$5.67 \times$}}
&\makecell{0.36\\\textcolor{myred}{$187.0 \times$}} \\
ViT 
&\makecell{18.23\\\textcolor{mygreen}{$3.09 \times$}} 
&\makecell{14.74\\\textcolor{mygreen}{$\textbf{3.83} \times$}} 
&\makecell{16.07\\\textcolor{mygreen}{$3.51 \times$}} 
&\makecell{56.41\\\textcolor{mygreen}{$1.00 \times$}} 
&-\tnote{3}
&\makecell{0.59\\\textcolor{myred}{$95.44 \times$}} \\
\bottomrule
\end{tabularx}
\end{threeparttable}
\footnotesize
\begin{tablenotes}
\fontsize{7}{8}\selectfont
    \item{1}. \textcolor{mygreen}{Secure} baseline for privacy-preserving and verifiable DNN training.
    \item{2}. \textcolor{myred}{Insecure} baselines without any protection. 
    \item{3}. The DarKnight prototype does not support Transformer-based model.
\end{tablenotes}
\label{tab:speedup}
\end{table}

\subsection{Inference Performance (\textbf{\req{1}} - Part II)}

\begin{figure}
  \centering
  \includegraphics[width=0.9\linewidth]{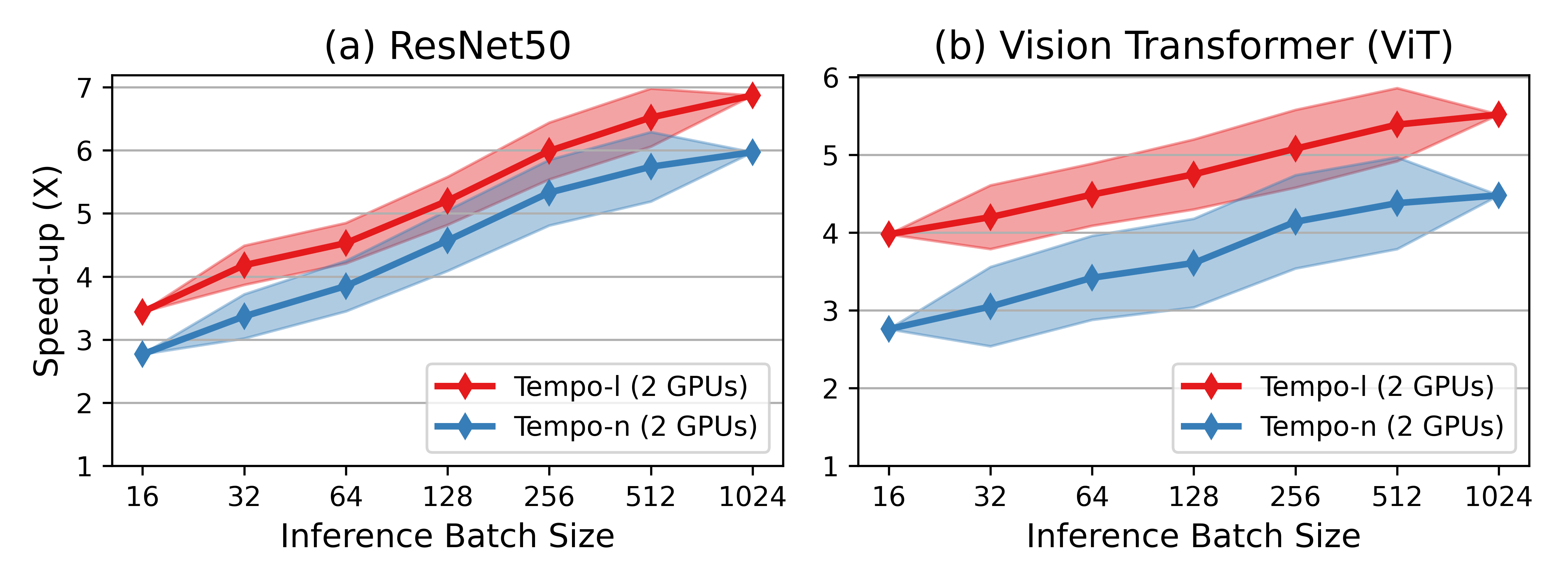}
  \vspace{-0.4cm}
  \caption{Inference speed-up of \name{} for two DNNs: (a) ResNet50 (b) ViT, compared with the secure baseline. The inference batch size is the number of images (in CIFAR-10 test set) fed into the DNN simultaneously.}
  \label{fig:exp-inf}
\end{figure}

For DNN inference tests, we choose MLCapsule~\cite{hanzlik2021mlcapsule}, a state-of-the-art privacy-preserving inference system, as the secure baseline.
We evaluate the inference speed-up of \name{} and the results are shown in~\autoref{fig:exp-inf}. Similar to training, we test the prototypes with batches of images as input and the time cost is averaged to a single image (for both \name{} and the baseline). For both prototypes and DNNs, we notice that performance improves when larger batch sizes are used.

\begin{figure}
  \centering
  \includegraphics[width=0.9\linewidth]{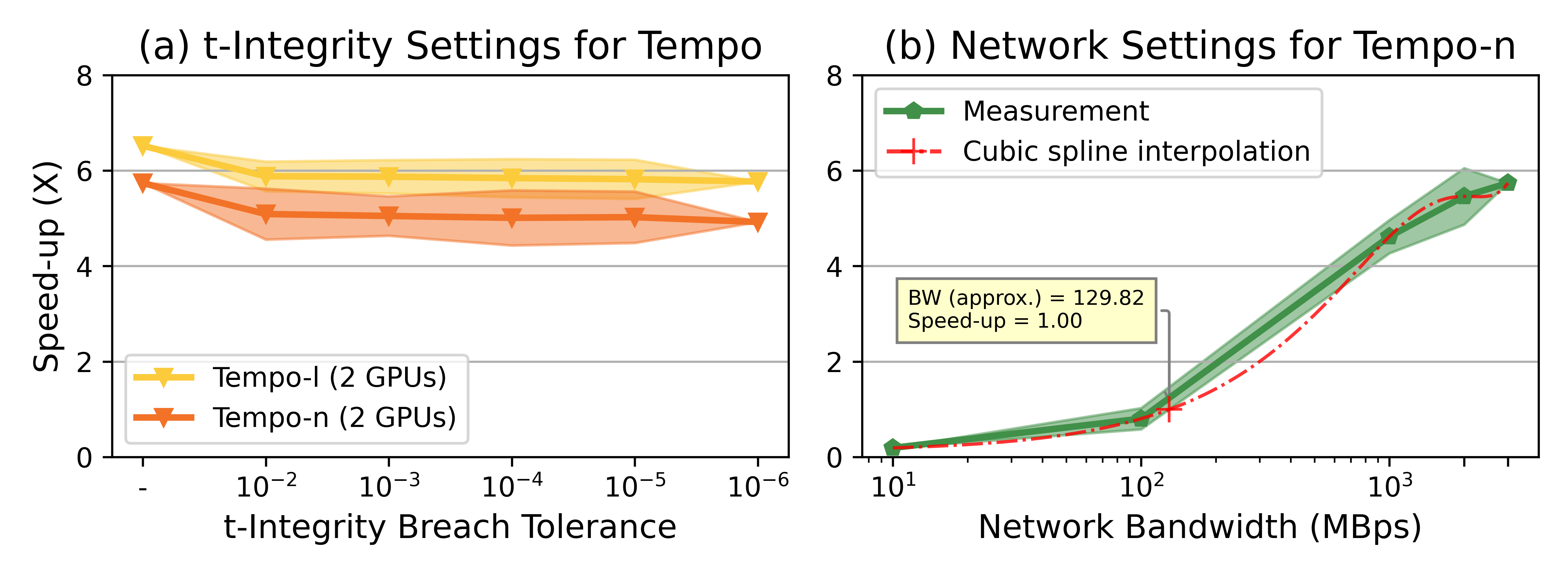}
  \vspace{-0.3cm}
  \caption{Inference speed-up of \name{} for ResNet50 compared with the secure baseline under different settings: (a) $t$-integrity tolerance (b) network performance. We set the inference batch size $= 512$.}
  \label{fig:exp-inf-veri}
\end{figure}

We also conduct evaluations for inference efficiency under different system configurations. 
\autoref{fig:exp-inf-veri} demonstrates the inference performance influenced by the integrity and network settings.
Recall in~\autoref{alg:decryption}, \name{} verify the integrity of the offloaded matrix multiplication using the Freivalds' algorithm~\cite{freivalds2005fast}. To achieve different levels of $t$-integrity, the result verification frequency may vary. 
For smaller error tolerance, more frequent verification is required. 
But experiments show that these extra calculations \emph{do not} lead to significant performance degradation (see \autoref{fig:exp-inf-veri} (a)).
We also investigate the performance of \name{} under different network settings. We evaluate \emph{\name{}-n} (2-GPUs) for various network bandwidths from \SI{10}{MBps} to \SI{3}{GBps} (see \autoref{fig:exp-inf-veri} (b)). Performance improves drastically with greater bandwidth. 
Using cubic spline to interpolate the measured data points, we infer that \emph{\name{}-n} (2-GPUs) losses its practicality when the bandwidth is smaller than \SI{129.82}{MBps}. By default, we use the largest possible bandwidth for all prior experiments on \emph{\name{}-n} prototype.

\subsection{Information Leakage of Real-world Attacks (\textbf{\req{3}})}

To further assess the robustness of \name{} and whether the theoretical privacy guarantee presents in~\autoref{sec:analysis} is effective in practice, we conduct model theft attacks~\cite{zhou2020dast}, \aka, model stealing, to steal the target model by querying it using \emph{synthetic inputs} since the real data for training the target model is normally unobtainable for the adversary.
We launch the DaST~\cite{zhou2020dast} attack to train a \emph{substitute model $I$} to imitate the performance of a trained ResNet50 model (dubbed the \emph{attacked model $T$}) inside \name{}. DaST feeds the 100$\%$ synthetic samples $\mathbf{\hat{X}}$ to $T$ and collects its predictions $T(\mathbf{\hat{X}})$ for substitute learning. 

We select a ResNet50 and initialized it with the \textbf{encrypted weights} $\mathbf{W}'$ copied from $T$ for the backbone of $I$.
For comparison, we use the same strategy to train another substitute model $I_o$, which is another ResNet50 initialized with \textbf{random weights}. 
Both models are trained using the same set of query pairs $\mathcal{Q}=\{\mathbf{\hat{X}},T(\mathbf{\hat{X}})\}$ collected by querying $T$.
$I$ and $I_o$ are then tested over $1,000$ images in the test set of CIFAR-10, yielding an accuracy of $25.91\%$ and $30.23\%$, respectively, while $T$ enjoys an accuracy of $90.37\%$ over the same set of images. 
This implies when suffering from model theft attacks, (a) the leakage of \name{} is modest; and more importantly, (b) the attacker's view of \name{} will not gain him any advantage.
It also noteworthy that \name{} and all similar TEE-assisted secure DL systems \emph{are not naturally defensible} against black-box attacks based on inference APIs~\cite{tramer2016stealing,papernot2017practical}.
Considering that \emph{training} on the cloud \emph{limits} the attacker from launching a large number of queries (\ie, $\left| \mathcal{Q} \right|$ is small), thus this type of attack only makes sense in terms of inference.
Nonetheless, by cooperating with other efforts~\cite{juuti2019prada}, \name{} can be used to further narrow down the attack surface.
\section{Related Work}
\label{sec:related}

\subsection{TEE for Privacy-preserving Learning}

Researchers have well explored privacy-preserving inference (\eg, \cite{tramer2018slalom, hanzlik2021mlcapsule}) and training (\eg, \cite{hashemi2021darknight, ng2021goten, niu20213legrace}) \wrt{} input privacy using both TEE and GPU. 
Slalom~\cite{tramer2018slalom}, an inference-only system that protects client input privacy, achieves its goal by using $\linear{\mathbf{W}}{\mathbf{X}} = \linear{\mathbf{W}}{(\mathbf{X}+\mathbf{r})} - \linear{\mathbf{W}}{\mathbf{r}}$ (they use $(\mathbf{X}+\mathbf{r})$ to blind the input $\mathbf{X}$). 
Since the weights $\mathbf{W}$ are kept in plaintext and the pre-computed factors $\linear{\mathbf{W}}{\mathbf{r}}$ make it impossible to update the weights during training, their method cannot be applied for both model privacy protection and training.
DarKnight~\cite{hashemi2021darknight} preserves the input privacy in both training and inference using a simpler matrix encoding $\linear{\mathbf{W}}{\mathbf{X}} \times \mathbf{A} = \linear{\mathbf{W}}{\mathbf{X}\mathbf{A}}$, with $\mathbf{A}$ the encoding matrix. Their approach is not feasible for protecting the model since $\mathbf{W}$ is not encoded (obfuscated) and is shared with all GPUs. 
Till recently, systems that protect the model privacy (\eg, \cite{zhang2021citadel, hou2021model,shen2022soter}) have received little attention, in which~\cite{hou2021model,shen2022soter} are proposed for inference and~\cite{zhang2021citadel} does not utilize GPUs.
SOTER~\cite{shen2022soter} protects both the model and input confidentiality based on the associative property of the linear operations in DNN, \ie, $\mu \linear{\mathbf{W}}{\mathbf{X}}=\linear{\mu\mathbf{W}}{\mathbf{X}}$, with $\mu$ a scalar blinding coin. However, SOTER fails short to cooperate DNN training\footnote{One possible reason is that the privacy-preserving ability from a scalar coefficient is weaker than matrices, the blinding coin $\mu$ needs to be refreshed more frequently to counter adversaries, making it less realistic in training.}.
Moreover, few work investigates GPU TEE for secure DL (\eg, \cite{volos2018graviton,hunt2020telekine}), requiring customized accelerators instead of employing publicly available (CPU) TEE and GPUs.

\subsection{Other Privacy-preserving Learning Methods}

Apart from TEE-based systems, a broader community has made substantial progress in using a variety of technologies to protect privacy in DL.
We mark some notable literature in this section, though, most of these methods aim for \emph{orthogonal} privacy-preserving goals with ours.
\emph{Homomorphic Encryption (HE)}~\cite{gilad2016cryptonets,jiang2018secure} applies a DNN to the encrypted input directly to make encrypted predictions. They exhibit theoretical privacy guarantees over the input but also incur multiple orders of magnitude slow-down, thus not suitable for large DNN training.
\emph{Secure Multi-Party Computation (MPC)}~\cite{mohassel2017secureml} is applicable for distributed learning when a master node is absent. Their main purpose is to protect the input and suffer from significant communication overhead as the number of participants increases.
\emph{Differential Privacy (DP)}~\cite{abadi2016deep} considers the training sets crowdsourced. They prevent an adversary from extracting parts of the training data of an individual user's sensitive information.
\emph{Federated Learning (FL)}~\cite{mcmahan2017communication} regards the data as decentralized, allowing data to be analyzed on individual devices to train a global \emph{shared} model. Researchers have combine FL, MPC, and DP (\eg, \cite{wei2020federated, kaissis2020secure, so2023securing}) to preserve an individual's input privacy. Defensive mechanisms against privacy breaches, such as deep leakage from gradients (DLG), have stimulated research in privacy-preserving FL community recently to protect the privacy of raw data~\cite{wang2022protect, gao2023automatic}.

\section{Concluding Remarks}
\label{sec:conclusion}

This paper presents \name{}, a distributed DNN training system designed to safeguard model privacy in cloud environments.
Leveraging TEE, \name{} guarantees both input and model confidentiality, while harnessing GPU for enhanced performance. This system integrates an obfuscation algorithm tailored for efficient training.
\name{}'s flexibility allows compatibility with various TEEs.
Evaluation findings validate \name{} as a robust solution, demonstrating its potential to fortify the privacy infrastructure within the MLaaS paradigm.

\newpage
\bibliographystyle{IEEEtran}
\bibliography{IEEEabrv,ref}

\end{document}